\documentclass[%
 reprint,
 amsmath,amssymb,
 aps,
]{revtex4-2}
\usepackage{xcolor}

\usepackage{graphicx}% Include figure files
\usepackage{dcolumn}% Align table columns on decimal point
\usepackage{bm}% bold math
\begin{document}

%\preprint{APS/123-QED}

\title{Decoherence and turbulence sources in a long laser}

\author{Amy Roche and Svetlana Slepneva}
\affiliation{
 Department of Physical Sciences, Munster Technological University, Cork, Ireland.
}
\author{Anton Kovalev}
\affiliation{ITMO University, Saint Petersburg, Russia}
\author{Alexander Pimenov and Andrei G. Vladimirov}
\affiliation{Weierstrass Institute, Mohrenstr. 39, 10117 Berlin, Germany}
\author{Mathias Marconi, Massimo Giudici and Guillaume Huyet}
\affiliation{Universit\'e C\^ote d'Azur, CNRS, INPHYNI, France}

%\affiliation{%

%\author{Anton Kovalev}
%\affiliation{
% This line break forced% with \\}%
%\affiliation{
% Third institution, the second for Charlie Author}%
%\affiliation{%
 %This line break forced with \textbackslash\textbackslash}%

\date{\today}% It is always \today, today,

\begin{abstract}
We investigate the turn-on process in a laser cavity where the roundtrip time is several orders of magnitude greater than the active medium timescales. In this long delay limit the electromagnetic field build-up can be mapped experimentally roundtrip after roundtrip. We show how coherence settles down starting from a stochastic initial condition.  In the early stages of the turn-on, we show that power drop-outs emerge, persist for several round-trips and seed dark solitons. These latter structures exhibit a chaotic dynamics and emit radiation that can lead to an overall turbulent dynamics depending on the cavity dispersion. 
\end{abstract}

%https://fr.overleaf.com/project/6352a6a7e7f5bf2bda2085cf
\maketitle

%\tableofcontents

The coherence property of light has fascinated researchers for several centuries and, following the laser discovery, numerous devices have exploited coherence for applications in industries, ranging from communication to metrology. Examples include Mach-Zehnder modulators for optical communications, swept-sources lasers for optical coherence tomography~\cite{fujimoto} or optical frequency combs for high precision metrology~\cite{Haensch}.

As for many of the laser characteristics, the theoretical analysis of the coherence property of single mode lasers, as a function of the pump parameter, is well-established~\cite{loudon2000quantum}. Below threshold, the photon statistics is well-described by a thermal distribution and the coherence time is inversely proportional to the gain bandwidth. Above threshold, the photon statistics is described by a Poisson statistics and the coherence time is inversely proportional to the Schawlow and Townes linewidth~\cite{schawlow1958infrared}.  Such studies have been the subject of recent interest in the context of nano-lasers where nontrivial photon statistics can emerge~\cite{wang2020superthermal} and high spontaneous emission factors can be engineered, giving rise for example to thresholdless lasers~\cite{rice1994photon}.   

The coherence property of multimode lasers has also been investigated in the context of the appearance of instabilities and the development of optical turbulence. These studies focused on the formation of optical vortices and the development of defect-mediated turbulence in the transverse section of wide aperture lasers \cite{arecchi1991vortices}.  Similarly, turbulence has been observed in the longitudinal dimension of the laser resonator where  coherent structures such as dark solitons~\cite{turitsyna2013laminar,tang2014dark,churkin2015stochasticity} or Nozaki-Bekki holes~\cite{slepneva2019convective} have been studied.

Coherent structures and turbulence have been thoroughly investigated in  long lasers operating above threshold \cite{turitsyna2013laminar,yarutkina2013numerical}, however there is still little information available about the transition to turbulence in these lasers in the course of the turn-on transient  when the pump parameter suddenly brings the laser above threshold. The analysis of the turn-on transient of lasers has already been a powerful tool to measure the lasers internal time scales.  
Early studies have, for instance, established that the turn-on transient of lasers relaxes to steady-state as an overdamped or a weekly damped oscillator, or even remains chaotic depending on the relative values of the photon, population inversion and polarization decay  times~\cite{tredicce1985}. It was also found that noise may alter the switching time \cite{balle1991statistics}.  While there are numerous experimental and theoretical studies about the dynamical and statistical properties of single mode lasers ~\cite{arecchi1967PRL, PRLstatistics,NATUREstatistics},  the evolution of the dynamical and statistical properties of multimode lasers, that can reach a turbulent regime, remains to be fully investigated.

%Recent studies of photon statistics in nano-lasers near threshold have shown that the statistics is closely connected to the emission of short pulses \cite{wang2020superthermal}.

%The laser response, when the laser is abruptly turned above threshold, has also been the subject of numerous studies starting soon after the laser discovery \cite{arecchi1967statistics}, with special attention given to the laser output power. 

%that lasers may reach their steady-state like an over damped or a weekly damped oscillator, or even remain chaotic depending on the relative values of the photon, population inversion and polarization decay  times. It was also found that noise may alter the switching time \cite{balle1991statistics}.
%While most of these studies provide valuable information on the evolution of the mean output power, there is little knowledge about the evolution of the photon statistic and coherence build-up during the turn-on transient of laser {\bf maybe we should cite the paper of Tito as it is in a turn-on transient}. 
%\textcolor{red}{}
%The main challenge for such a study is that these properties will evolve quickly during the first few cavity round trip times, but this can be overcome with the aid of a long cavity laser. In addition, by using a localized gain medium, it is possible to store and analyse the evolution of the light at each round trip-time.

\begin{figure}[h!]
\begin{center}
\includegraphics[height=3.1cm]{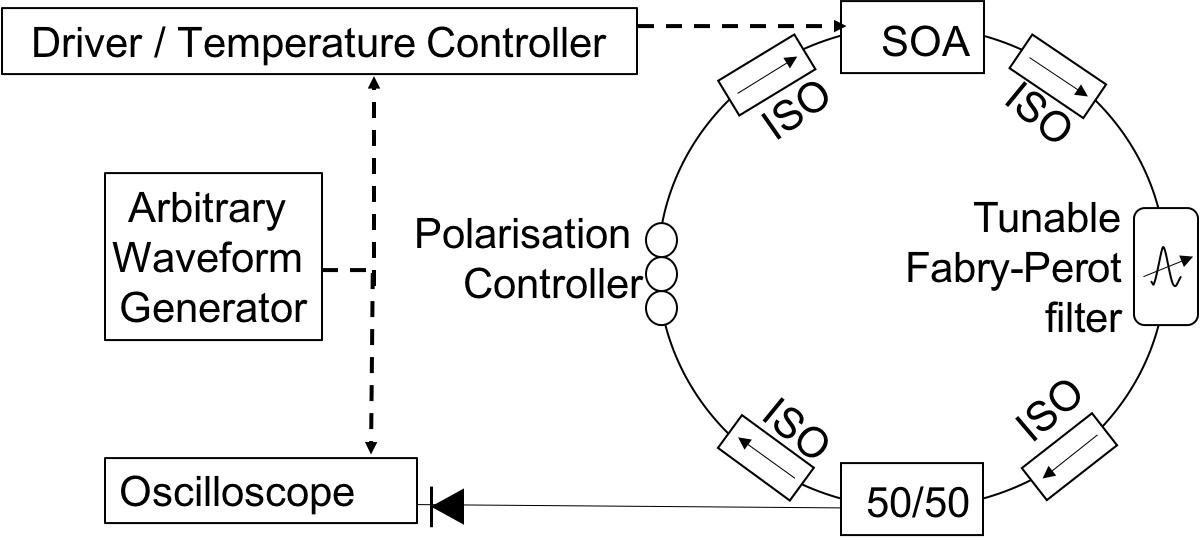}
\end{center}
\caption{Experimental setup of the semiconductor laser with a ring fibre cavity. SOA: semiconductor optical amplifier. ISO: fibre optical isolator. Half of the laser power is used for the detection.} 
\label{fig:setup}
\end{figure}

This paper aims to provide an experimental and theoretical description of the evolution of the dynamical and statistical properties of light during the turn-on transient of a highly multimode laser. By using a long fiber ring cavity containing a short semiconductor gain medium, we are able to record and analyze the  phase and intensity variations, and reconstruct the evolution of the photon fluctuations and coherence within each round trip.   This study demonstrates that the transition from thermal to Poisson statistics can be described using a one-dimensional map of the laser intensity which reveals the appearance of power drop-outs during the first few round trips. These power dropouts slowly disappear and the photon statistics relaxes towards a Poisson statistics. However they may also seed holes similar to dark solitons or Nozaki-Bekki holes \cite{nozaki1984exact}. In the normal dispersion regime, these holes travel with a constant speed while their cores display a chaotic dynamics that leads to the radiation of decaying turbulent bursts that travel faster than the holes. 
In the anomalous dispersion regimes, turbulence bursts emitted by a hole can reach and alter the dynamics of the preceding hole and, as a result lead to a chaotic trajectory of the holes.  This interaction between holes, which is similar to defect-mediated turbulence, constitutes the main source for coherence degradation in long lasers operating in the anomalous dispersion regime. 

%In addition, as opposed to single mode lasers where the stationary regime above threshold is always Poissonian \cite{arecchi1967statistics}, our study of a long lasers coherence build up allows to connect early round trip statistics to diverse post-turn-on states, characterized by large intensity and phase fluctuations within the round trip which are induced by the highly multimode dynamics of the system \cite{gowda2020turbulent}. We present a new mechanism of coherence degradation induced by the radiation of localized structures such as dark solitons connecting the off and the lasing state, which are formed during the transient of the laser turn-on as we know dark solitons are interconnections of different (nontrivial) modes.
%{\bf maybe we should link the two above paragraph since the coherence time is degraded due to the presence of dark solitons}

\begin{figure}[t!]
\begin{center}
\includegraphics[height=5.5cm]{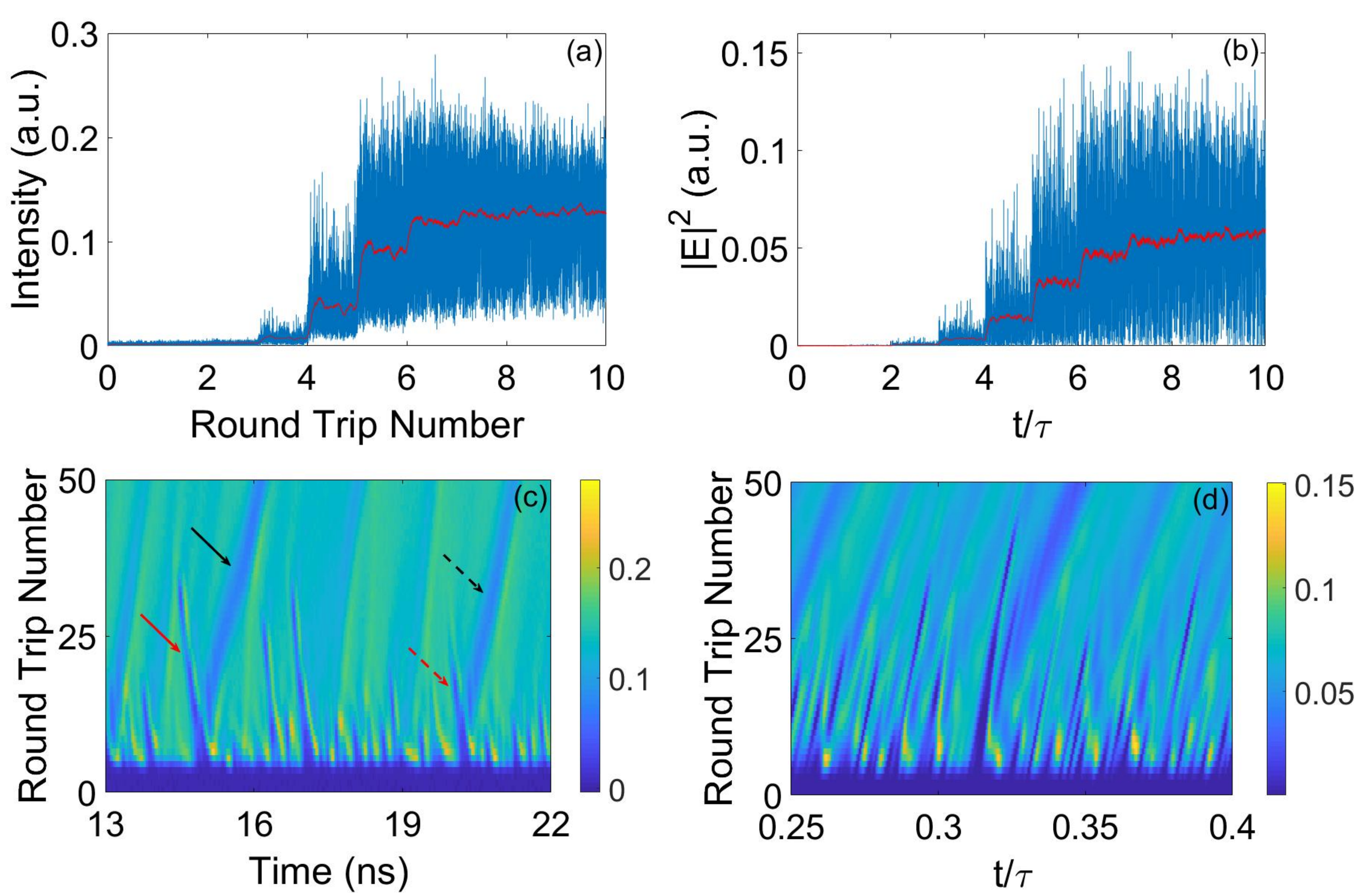}
\end{center}
\caption{(a) Experimental time trace of the intensity over the first 10 round trips after the turn on of the SOA (round trip 0) recorded at 12 GHz bandwidth (blue) and numerically filtered down to 100 MHz (red). (b) The simulated time traces of the laser turn-on process over the first 10 round trips (blue) and the numerically filtered down to 100 MHz intensity signal (red). (c) Experimental 2D map of a 9~ns zoom of the laser intensity time trace over the first 50 round trips. The arrows indicate two distinct types of power drop-out structures. (d) Simulated intensity evolution of the first 50 round trips.} 
\label{fig:start}
\end{figure}

The experimental setup is shown in Fig.~\ref{fig:setup}. The laser includes a semiconductor optical amplifier (SOA) with a gain peak near $1.3~\mu m$ and a tunable optical bandpass filter with $10$~GHz transmission bandwidth. 
The total cavity length of about $21$~m, corresponding to $104.32$~ns cavity roundtrip time was mostly comprised by the added lengths of the fiber pigtails of all the cavity components. In order to analyse the turn on dynamics, the bias current $I_b$ of the  SOA was quickly turned on to the values of $1.2$ or $1.6$ times of its threshold current $I_{th}=92$~mA. 
%At this value the laser generated one NBH per roundtrip~\cite{gowda2020turbulent} {\bf should we speak about NBH here??}. \sout{For the given parameters, $I_{th}$ was $92$~mA.} 
The driver is quickly turned on by a square-shaped signal having a $20$~ns rise time delivered by an arbitrary waveform generator.  This time is much shorter than the laser cavity roundtrip time. A $50/50$ fiber splitter placed after the filter is used to extract the light out of the cavity, which is detected and monitored by a real time oscilloscope with 12~GHz bandwidth, synchronously triggered by the same waveform generator. Using the interferometric technique described in \cite{Kelleher2010}, with the aid of a stable narrow linewidth laser, both the relative phase and the amplitude of the electric field of the laser are recorded in a single shot measurement, allowing access to the evolution of the dynamics within each round trip of the laser transient.

\begin{figure}[t!]
\begin{center}
\includegraphics[height=2.8cm]{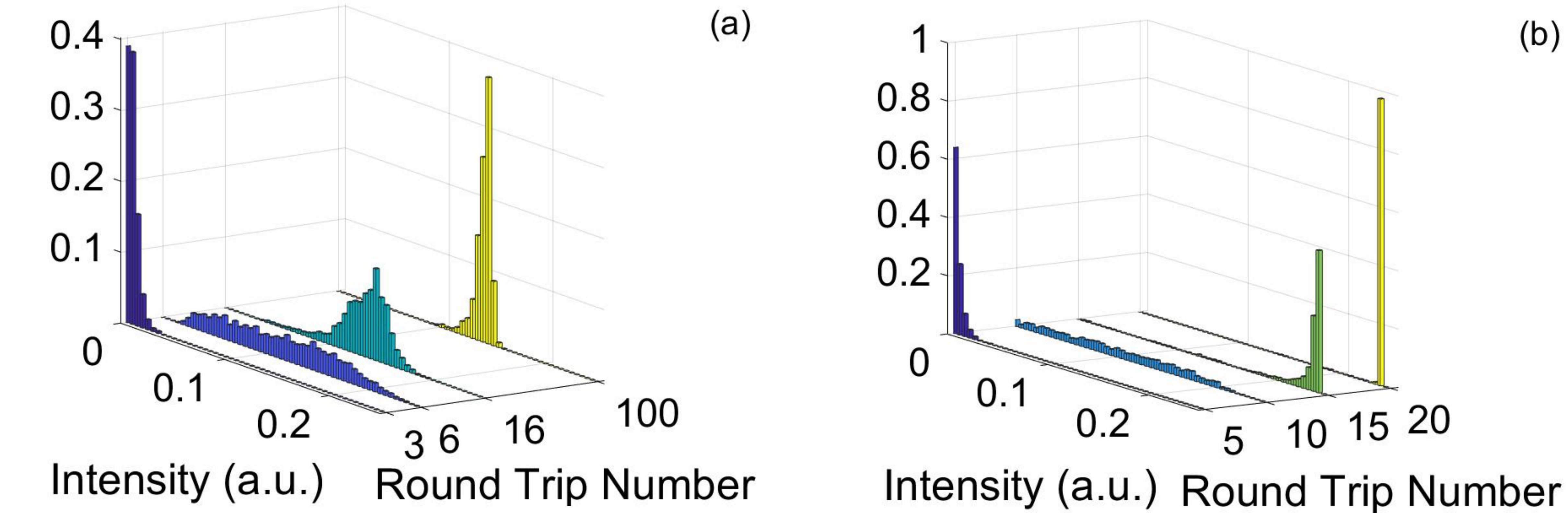}	
\end{center}
\caption{Experimental (a) and numerical (b) probability distributions of the laser intensity for successive roundtrips during the laser turn-on.  Numerical histograms are obtained with the 1D intensity map.}
\label{fig:hist}
\end{figure}

The experimental data were modeled using the delay differential equation (DDE) model as in~\cite{Pimenov17}

%\begin{align}
%& \gamma^{-1}\dfrac{dE(t)}{dt} + (1 - i\Delta )E(t) = A(t-\tau)\left[E(t-\tau)+ \epsilon P(t-\tau) \right]+\beta(t),  \label{eq:field} \\
%& \dfrac{dP(t)}{dt} =-(\Gamma -i\Omega ) P + E(t),\\
%&\gamma _{g}^{-1}\dfrac{dG(t)}{dt} = g_{0}-G(t)-(e^{G(t)}-1)\left\vert
%E(t)+ e^{i \varphi} P(t)\right\vert ^{2}  \label{eq:gain}
%\label{eq:1}
%\end{align}

\begin{eqnarray}
\gamma^{-1}\dfrac{dE(t)}{dt} &+& (1 - i\Delta )E(t)= \nonumber\\&+&A(t-\tau)\left[E(t-\tau)+ \epsilon P(t-\tau) \right]+\beta(t),\label{eq:field} \\
\dfrac{dP(t)}{dt} &=&-(\Gamma -i\Omega ) P - E(t),\\
\gamma _{g}^{-1}\dfrac{dG(t)}{dt} &=& g_{0}-G(t)-(e^{G(t)}-1)\left\vert
E(t)+ \epsilon P(t)\right\vert ^{2} \nonumber\\ \label{eq:gain}
\end{eqnarray}

with $$A(t)= \sqrt{\kappa} \exp ( (1-i \alpha ) G(t-\tau)/2 )e^{i\varphi},$$ where  $t$ is the time, $\tau$ is the cold cavity round trip time,
$E(t)$ is the complex electric field envelope, $G(t)$ is the gain,  $P(t)$ is the polarisation of the optical fibre which can be tuned from normal to anomalous dispersion by changing the sign of $\Omega$.  $\kappa $ describes the linear cavity losses, $\gamma _{g}$ is the carrier density decay rate, $g_{0}$ is the pump parameter, $\alpha $ is the linewidth enhancement factor, $\gamma $ is the
filter bandwidth, $\Delta$ is the central frequency of the filter, and $\beta (t)$ is a white noise source with a variance of $3.2*10^{-3}$ during the first round trip, $t<\tau$, and $0$ afterwards. $\Gamma$ and $\Omega$ is the polarization decay rate and frequency frequency detuning. The parameter values used in the simulations were: $\tau = 104.32$~ns, $\gamma =
19$~GHz, $\gamma _{g}^{-1} = 1$~ns, $\kappa = 0.036$ (corresponding to $-14.5$~
dB losses), $\alpha = 3.5$. 
%The variance value $\delta_{11}=\delta_{22}=3.2\times10^{-3}$ was chosen to match the experimental time of build-up process. 
%\andrei{$\delta_{11}$ and $\delta_{22}$ ARE NOT INTRODUCED ABOVE} 

The
filter was stationary, and $\Delta(t) = 0$. The pump parameter threshold
value was taken as $g_{0,thr}=-\ln\kappa$. The model (\ref{eq:field})--(\ref{eq:gain}) was numerically integrated by
means of an Euler scheme with a step $dt = 1.05$ ps. The
initial condition corresponded to the laser-off state $E(t<0)=0$, $G(t<0)=g_0$, and the laser was turned on at the moment $t=0$.

 The upper panels of Fig.~\ref{fig:start} depict the first 10 round trips of the laser turn-on transient observed experimentally (a) and calculated numerically (b). Spontaneous emission, emitted by the SOA as soon as the pump current increases, is re-amplified after each round trip. As a result,  the laser intensity, averaged over one round-trip, increases in a step-wise fashion and reaches its steady-state value after a few round trips.  The laser continues to exhibit large amplitude fluctuations for a much longer duration and, to obtain further insight on this  dynamics, we represent the evolution of a 20 ns time span over 50 round trips on the two lower panels. These intensity maps reveal the existence of two types of power drop-out structures. The first one appears within the first round-trips of the turn-on transient and disappear after about 25 round trips (red arrows on the panel).
The second type of power drop-outs remains for a much longer round trip number and have a larger round-trip  time than the first type of drop-outs (black arrows on the panel). Concretely, this means that the first type of drop-outs have a larger group velocity than the second type.

%The numerical simulations of the equations (\ref{eq:field})-(\ref{eq:gain}) also reveal the existence of power drop-outs in the initial stages of the turn-on transient.   This behaviour, shown on Fig.~\ref{fig:start},
%can be explained by reducing equation~(\ref{eq:field}) to a one dimensional map in the case of zero dispersion, $\epsilon=0$, and infinite filter bandwidth, $\gamma^{-1}=0$. This map, which is obtained by neglecting the time derivative in Eq. (\ref{eq:field}) and adiabatically eliminating the carrier density  $G$ from Eq. (\ref{eq:gain}) in the limit $G\ll 1$, reads
%\begin{align}
%p_n = \kappa \exp \left(\frac{g_0}{1+p_{n-1}} \right) p_{n-1},\label{map}
%\end{align}
%where  $p_n=p(n\tau)$. When the laser is biased below threshold, this map has only one fixed point $p=0$. Above threshold, the solution $p=0$ becomes an unstable fixed point and $p_{lasing}=-g_0/\ln{\kappa} -1$ becomes the new stable fixed point. We note that the trajectories that start with an initial condition in the vicinity of the unstable fixed point will converge slower towards the lasing steady-state than the one starting further from the unstable fixed point.  The difference in the convergence time of the trajectories is at the origin of the formation of power dropouts in the initial roundtrips. 

%Numerical simulations of the equations (\ref{eq:field})-(\ref{eq:gain}) also reveal the existence of power drop-outs in the initial stages of the turn-on transient.

Power dropouts, observed both experimentally and numerically in the initial stage of the transient, are a spatio-temporal representation of the stochastic nature of the switch-on dynamics. The turn-on dynamics of single mode lasers was experimentally observed~\cite{arecchi1967PRL} and theoretically investigated using the Fokker-Planck equation~\cite{Risken67}.  
In the case of long lasers, the filter bandwidth determines the coherence time of the emission during the first round trips, and, as a result, the recorded time series are well-described by thermal noise. After each round trip, the statistical fluctuations are modified as the laser intensity is re-amplified. To describe this behaviour, one can reduce the equations~(\ref{eq:field})-(\ref{eq:gain}) to a one dimensional map describing the evolution of the laser intensity after each round trip,
 $p_n=|E(n\tau)|^2$,
\begin{align}
p_n = \kappa \exp \left(\frac{g_0}{1+p_{n-1}} \right) p_{n-1}.\label{map}
\end{align}
The evolution of the transient probability distribution can be obtained iterating the mapping from a set of initial conditions following a thermal distribution. The evolution of the transient probability distribution functions (PDF)  of the laser intensity,  calculated from numerical and experimental data, are shown in Fig.~\ref{fig:hist}a.  These PDF's show a similar evolution as observed in \cite{arecchi1967PRL, Risken67}  as they display a  conventional thermal statistics for the first round-trips  before developing a broad bell-shaped  distribution. After a few round trips the PDF exhibits a sharp peak near the average laser intensity  but with a tailed distribution toward the low intensities due to the existence of power dropouts. In the case of the experiment presented here, the low intensity tail can be associated with power dropouts that are disappearing slowly.

\begin{figure}[t!]
\begin{center}
\includegraphics[height=5cm]{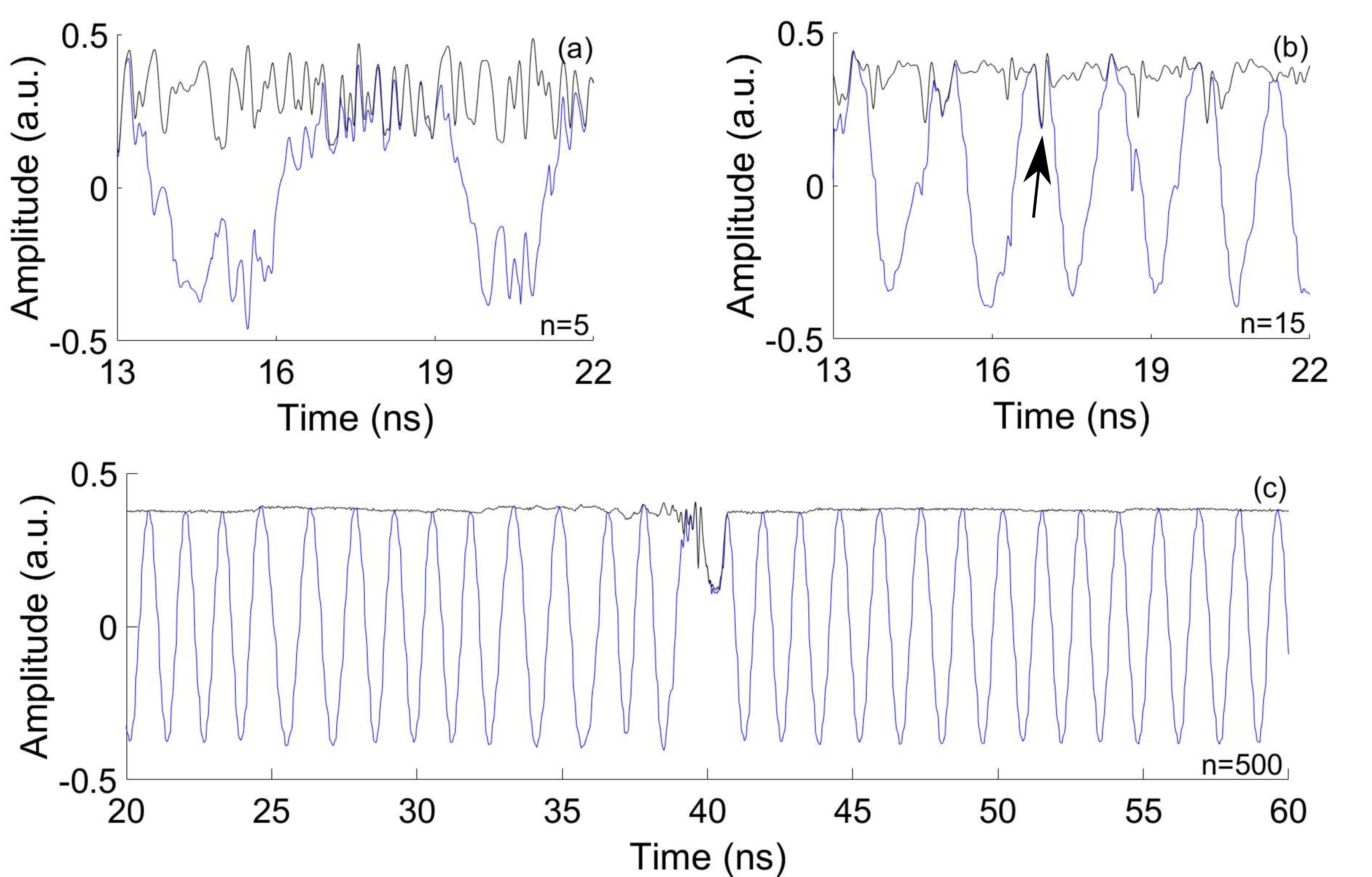}
\end{center}
\caption{
Experimental time trace of the laser electric field  (blue) and intensity (black) at round trips 5 (a), 15 (b), and 500 (c). The arrow in panel b indicates the phase discontinuity induced by a reminiscent holes observed during the first round trips.}
\label{fig:Efield}
\end{figure}

%To follow the evolution of the photon statistics in the turn on transient, we start with an ensemble of $N$ initial conditions $p^{(i)}_1$, with $1<i<N$, following a thermal distribution characteristics of spontaneous emission. The values of $p^{(i)}_n$ are calculated using equation (\ref{map}). The PDF's, calculated from these trajectories, are shown in Fig.~\ref{fig:start}(b). The PDF display the same low power tail as experimentally observed. This tail corresponds to sequences $p^{(i)}_n$ that converge slowly toward  $p_{lasing}$ as $n\rightarrow \infty$ as there are the closest to $p=0$ for $n=1$. These trajectories correspond to the connection from the non-lasing to the lasing state.

To gain further insight about the coherence build-up, we measured the temporal evolution of the electric field~\cite{Kelleher2010} during the turn on transient.
Fig.~\ref{fig:Efield} represents the experimental temporal evolution of the laser intensity and electric field  within a $20$ ns time window during the round-trip number 5, 15 and 500. During the $5^{th}$ round-trip, the intensity fluctuates near its mean value while the phase exhibits large and aperiodic variations. At roundtrip 15, the electric field displays a periodic evolution with small phase discontinuities near the reminiscent holes observed during the first round trips  (marked by the black arrow). 
At roundtrip number 500, we observe that the electric field tends toward a monochromatic evolution but some intensity power drop-outs induce phase discontinuities~\cite{gowda2020turbulent}. Measurements of the optical frequency before and after the power drop-outs demonstrates that these structures connect monochromatic waves with slightly different frequencies and are therefore the analog of sinks and sources of travelling waves observed in spatially-extended systems. They are commonly described within the framework of the complex Ginzburg-Landau equation~\cite{Aranson}.

%near such heteroclinic connections. Panel a depicts a 9ns time window during the $5^{th}$ round-trip where the two-dimension map of Fig.. exhibits numerous heteroclynic connections. For this round-trip number, the laser intensity displays small functuations around its mean value but phase discontinuities can be inferred from the temporal evolution of the electric field.  

%In the initial round trips (panel a) heteroclinic 
%connections are numerous and induce chaotic fluctuations of the electric field. After a few roundtrips, we observe that the phase of the laser has stabilized but some phase discontinuities are still induced by the  heteroclinic connections in the intensity buildup.

\begin{figure}[t!]
\begin{center}
\includegraphics[height=5.7cm]{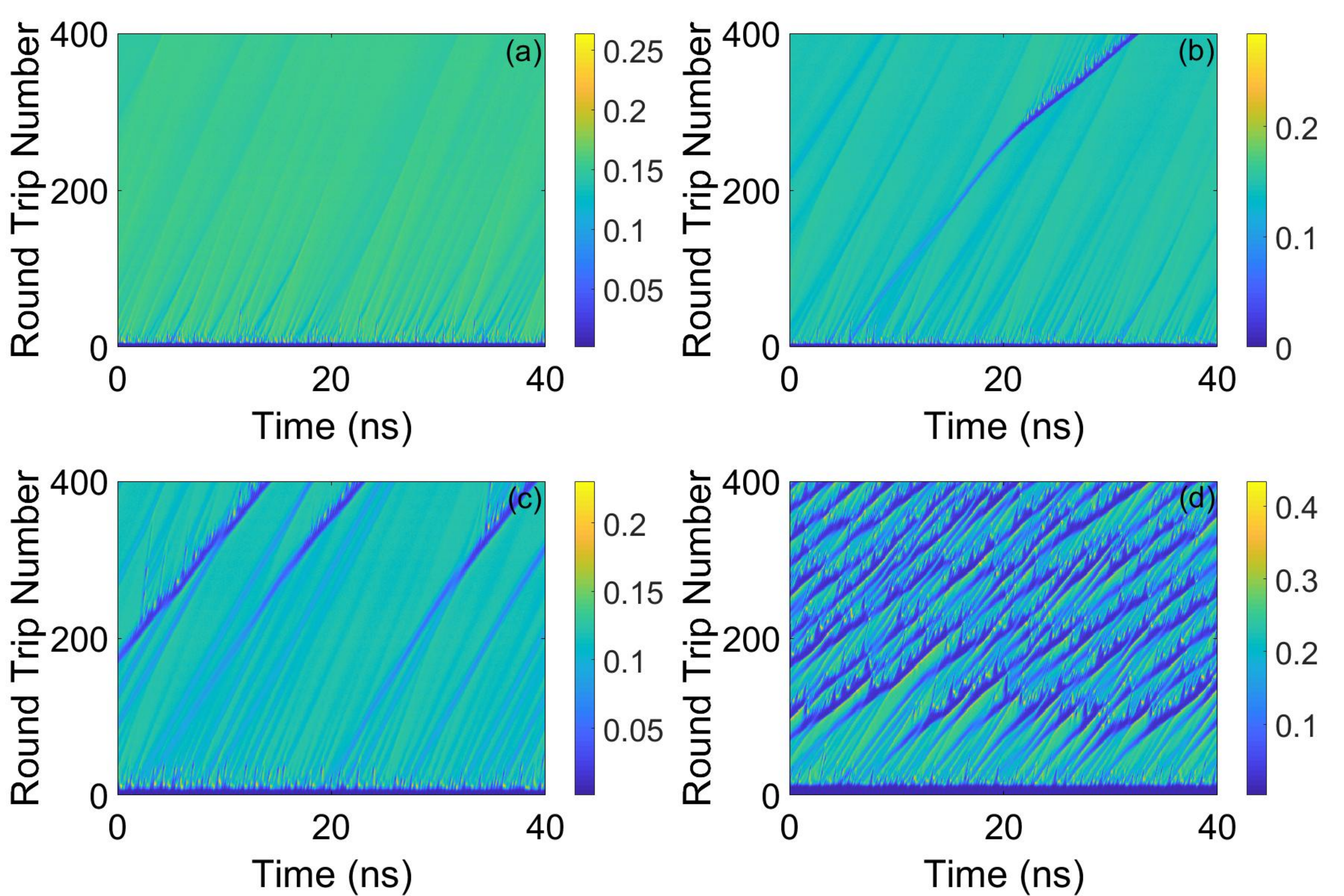}	
\end{center}
\caption{Maps of the laser intensity for different pump power and operating wavelength. a,b) Two different transient build-ups obtained for $P = 1.6*P_{th}$, operating at 1310 nm (normal dispersion). c) $P = 1.2*P_{th}$, filter at 1310 nm (normal dispersion). d) $P = 1.2*P_{th}$, filter at 1360 nm (anomalous dispersion).}
\label{fig:longtrace}
\end{figure}

\begin{figure}[h!]
\begin{center}
\includegraphics[height=5.1cm]{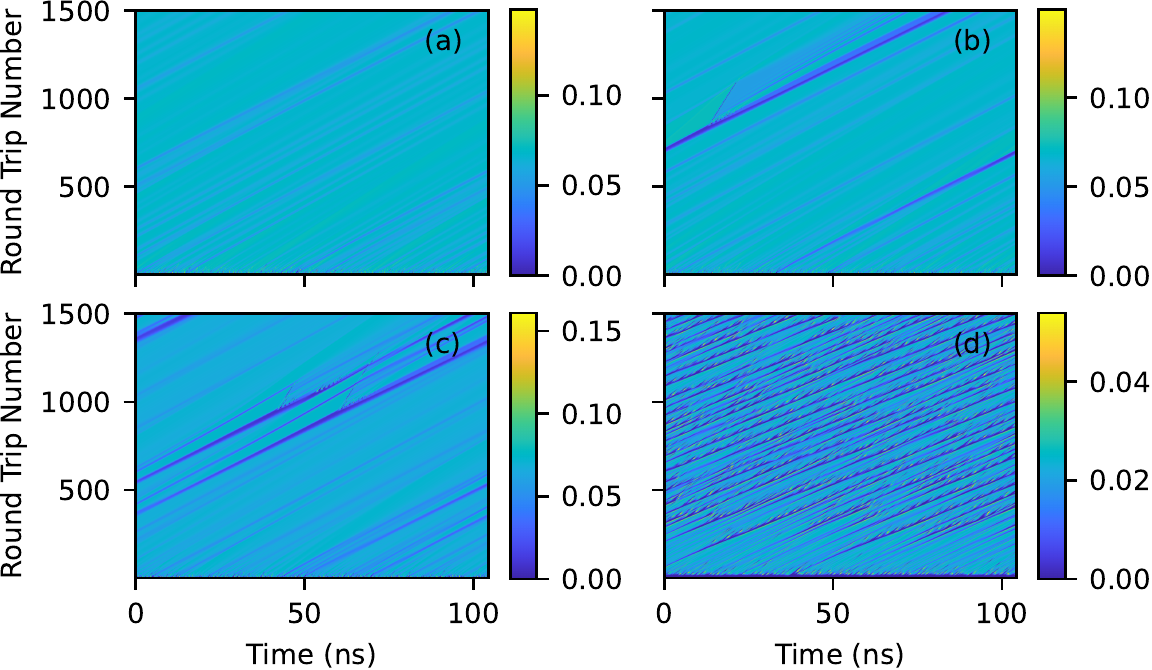}
\end{center}
\caption{Numerical maps of the laser obtained by numerical integration of Eqs. (\ref{eq:field})-(\ref{eq:gain}) to match the experimental data shown in Fig.~5. Transient build-ups calculated in zero dispersion regime ($\epsilon = 0$) with $g_0 = 1.6 g_{0,thr}$ (a), $g_0 = 1.2 g_{0,thr}$ (b,c), and anomalous dispersion regime (d) with $g_0 = 1.2 g_{0,thr}$, $\epsilon = 30.4$ GHz, $\Omega = -30.4$ GHz, and $\Gamma=15.2$ GHz. The other parameters are the same as in the main text.}
\label{fig:delta}
\end{figure}

\begin{figure}
\begin{center}
\includegraphics[height=3cm]{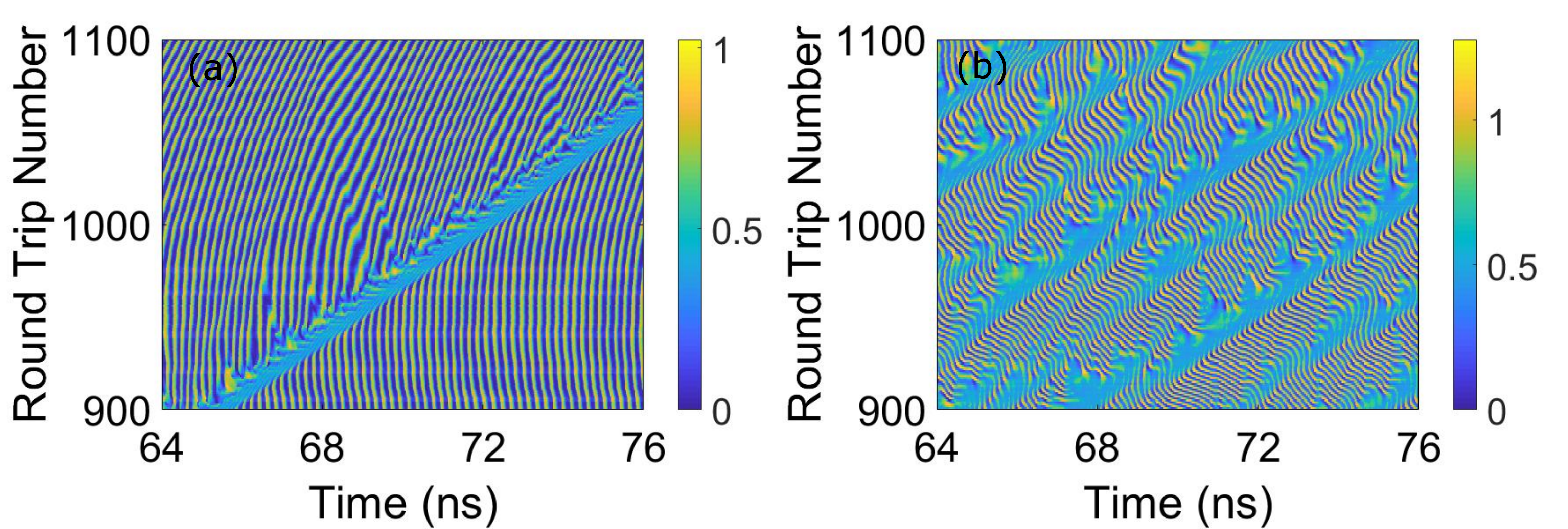}
\end{center}
\caption{2D map of the electric field for (a) 1310nm, showing a zoom of a single drop and (b) 1360nm, showing the interaction of multiple drops.}
\label{fig:mapEfield}
\end{figure}

The behaviour of these coherent structures depends on the dispersion, pump power and can also vary between each consecutive transients as illustrated in Fig.~\ref{fig:longtrace}.  The first two panels describe two different turn-on transients obtained for the same parameters, {\it i.e.} pump power at 1.6 times threshold and  filter central wavelength at $1310$ nm.  Some turn-on transients lead to the formations of domains with slightly different intensities that are connected by abrupt transitions. The measurement of the time-resolved laser frequency indicates that these domains have slightly different frequencies and the abrupt transitions can be associated with fronts connecting different travelling waves. For other transients, power drop-outs persist as shown on Fig.~\ref{fig:longtrace}b. We note that the group velocity of these power drop-outs varies during the transient and co-exist with small intensity variation shocks. For example, in panel b the group velocity abruptly decreases around roundtrip 300 when the dropout amplitude increases. This indicates that power dropouts observed in the few initial round-trips seed coherent structures that connect domains with different frequencies. In such a case, the group velocity should depend on various parameters including the amplitude of the dropout and the frequency difference between the domains that connect the dropout. However, we note that the group velocity of the power dropouts is smaller than the group velocity of the small intensity variation shocks.
In addition, the dropouts display a chaotic dynamics as described in~\cite{gowda2020turbulent} that leads to the emission of turbulent bursts from the holes as observed after roundtrip 300 in panel b. Fig.~\ref{fig:mapEfield} depicts the evolution of the electric field between round-trips 900 and 1100. In this figure, we observe the emission of turbulent bursts from a hole which separates two different domains. We note that the group velocity of these burst is larger than the group velocity of the holes. Taking into account only the linear effect of the spectral filtering on the additional delay time $\delta\tau$  per cavity round trip of the intensity perturbations near the lasing threshold, this time can be estimated as $\delta\tau\approx\gamma^{-1}$ \cite{vladimirov2005model}. Although this expression provides a rough approximation of the additional delay time of the dropouts, numerical simulations give smaller values of this time dependent on the pump parameter. %(see Supplementary material). 
This dependence may be attributed to the influence of the system nonlinearities on the delay time $\delta\tau$.
For lower injection current, we observe that more power drop-outs persist as described in Fig.~\ref{fig:longtrace}c. The dynamics of each power drop-out is similar to that observed at higher injection current, {\it i.e.} each dropout exhibits a chaotic dynamics and emits intensity bursts. Since small intensity variation shocks and power dropouts travel with different frequencies, they can collide as shown on panel c.  During the collision, the dropout group velocity is momentarily decreased but rapidly comes back to its original state after the shock. In this regime, intensity bursts emitted by chaotic holes do not interact with other structures as they decay rapidly as soon as they are ejected from the holes. 
The situation is drastically modified when the laser is operated in the anomalous dispersion regime (the intracavity filter is set at 1360 nm), as illustrated in Fig.~\ref{fig:longtrace}d) and Fig.~\ref{fig:mapEfield}d). In this regime, the number of power dropouts per round trip is much larger and the radiation emitted from each power dropout reaches and modifies the dynamics of the preceding dropout thus leading to a chaotic trajectory of the dropout core. Such a behaviour constitutes, to our knowledge, a new type of dynamics in defect-mediated turbulence ~\cite{coullet}.
To quantify the degree of decoherence induced by such dynamics, we compute the coherence time as a function of the roundtrip number for each situation  shown in Fig.~\ref{fig:longtrace}. The coherence time per roundtrip is obtained by integrating the optical power spectrum \cite{butlersingleshot}:
\begin{eqnarray}
\tau_c&=& \frac{\int_{-\infty}^{+\infty}|S(\nu)|^2d\nu}{(\int_{-\infty}^{+\infty}|S(\nu)|d\nu)^2}
\end{eqnarray}

\begin{figure}
\begin{center}
\includegraphics[height=6.2cm]{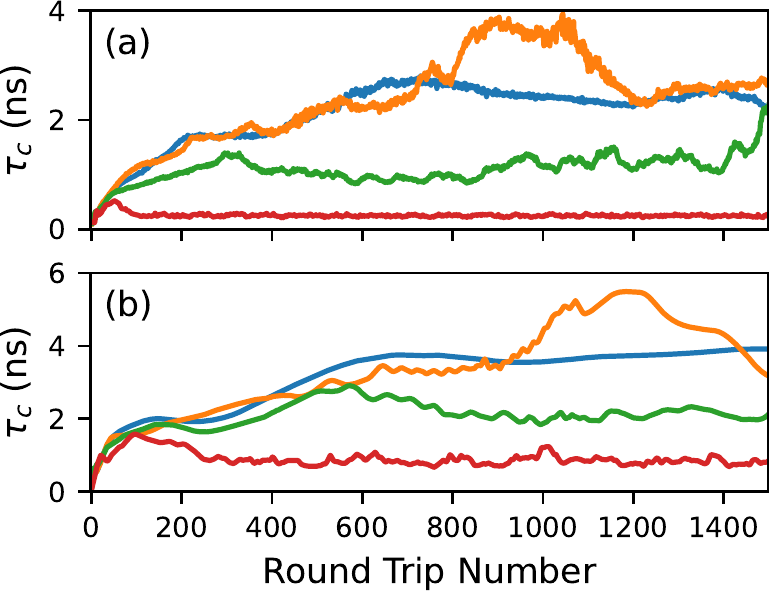}
\end{center}
\caption{Coherence times $\tau_c$ as a function of the round trip number. (a) Experiment: transient resulting in no drop (blue) corresponding to the situation shown in Fig.~\ref{fig:longtrace}(a), single drop  (orange) corresponding to the situation shown in Fig.~\ref{fig:longtrace}(b), multiple drops (green) corresponding to the situation shown in Fig.~\ref{fig:longtrace}(c), defect-mediated turbulence in the anomalous disperson regime (red) corresponding to the situation shown in Fig.~\ref{fig:longtrace}(d).
 (b) theory. Coherence times are calculated for simulated transients that identical to the experimental ones. 
 %(see Supplemental).  
 }
\label{fig:coherence}
\end{figure}

 We see from Figs.~\ref{fig:delta} and \ref{fig:coherence} that the model (\ref{eq:field})-(\ref{eq:gain}) is able to quantitatively retrieve the experimental results. In Fig.~\ref{fig:delta} numerical maps of laser intensity illustrating transient build-ups are shown in zero (a,b,c) and anomalous (d) dispersion regimes. They are in qualitative agreement with the experimental data shown in Fig.~\ref{fig:longtrace}.  Figure \ref{fig:coherence} shows that the coherence time quickly increases during the first 30 roundtrips, which confirms the observations made earlier. Afterwards we may distinguish between three types of behavior. In the case where the transient results in no drop or a single drops (blue and orange curves respectively), we observe that the coherence time converges to a value of about 3 nanoseconds after  approximately 500 roundrips.  In the multiple drop case (green curve), we notice a decrease in the coherence time with respect to the former cases. This can be explained by the collisions experienced by the multiple coherent structures present in the roundtrip. Finally, the situation is drastically modified in the anomalous dispersion regime. In fact, we observe that, after a rapid increase in the first roundtrips, the experimentally measured coherence time decreases and reaches the limit of about 100 ps corresponding to the inverse of the filter bandwidth. Although the dynamics of the laser is highly multimode in this situation, the coherence time  remains remarkably constant with further increase of the roundtrip number. 
%In a  This indicates that the photon statistic converges faster than the laser coherence. In addition, fig. 3 e shows that the heteroclinic connections induce some phase discontinuities {\bf can we say this} of the electric field.
%We note that during \andrei{......} the electric field displays a chaotic dynamics while the laser intensity converges towards its steady-state value. 

In conclusion, we have presented a roundtrip-resolved evolution of the intensity and phase decoherence during the turn-on of a long laser. Such lasers can exhibit a wide variety of temporal dynamics in the stationary regimes above threshold depending on the pumping power and group velocity dispersion. We have demonstrated that transition from thermal to Poisson statistics  occurs  after apporximately 10 roundtrips and is a temporally inhomogeneous process within the roundtrip. The intensity inhomogeneities are the source of phase discontinuities that can survive after multiple roundtrips and give rise to coherent structures or turbulent phenomena in anomalous dispersion regime. We have evidenced that the turbulent regime is mediated by the collisions between multiple phase defects that acts as a temporal decoherence mechanism.
%Funding
%\section*{Funding}

We acknowledge funding for the European Commussion for the projects H2020-MSCA-IF-2017, ICOFAS (800290) and H2020-MSCA-RISE-2018 HALT, from the Munster Technological University for funding of AR's PhD Scholarship 2018; Région PACA for the OPTIMAL project; Ministry of Science and Higher Education of the RF (Grant 2019-1442); Deutsche Forschungsgemeinschaft (DFG project No. 445430311).

%Disclosures
%\section*{Disclosures}
The authors declare no conflicts of interest.
%\section*{Supplementary material}
% Bibliography
\bibliography{ref}

%\bibliographyfullrefs{ref}

\end{document}